\documentclass[a4paper,10.5pt,twocolumn,twoside]{article}

\usepackage{graphicx}
\usepackage{color}
\usepackage[a4paper, left=1.4cm, right=1cm, top=1.7cm, bottom=1cm]{geometry}
\usepackage[super,comma,sort&compress,square]{natbib}
\usepackage{amsmath}
\usepackage{float}
\usepackage{latexsym}
\usepackage{amssymb}

\usepackage{fancyhdr}
\usepackage[utf8x]{inputenc}
\usepackage[T1]{fontenc}

\begin{document}
\pagestyle{fancy}
\def\headrulewidth{0.5pt}
\def\footrulewidth{0pt}
\lhead{Nanotechnology 26 (2015), p.~425301} 
\chead{}
\rhead{DOI: 10.1088/0957-4484/26/42/425301}

\lfoot{} 
\cfoot{}
\rfoot{}

\twocolumn[
  \begin{@twocolumnfalse}
  {\huge \bf Ordered FePdCu nanoisland arrays made by templated solid-state dewetting}

  \hspace{1.1cm}
  \parbox{.87\textwidth}{
    \vspace{4ex}
    \Large \textsf{M.~Krupinski, M.~Perzanowski, A.~Zarzycki, Y.~Zabila, M.~Marszałek}
    \vspace{1ex} \\
    \normalsize Institute of Nuclear Physics Polish Academy of Sciences, Deparment of Materials Science, Radzikowskiego 152, 31-342 Krakow, Poland 
    \vspace{1ex} \\
    \normalsize \text{email: Michal.Krupinski@ifj.edu.pl}

    \vspace{2ex} 
    \noindent
     \textbf{Abstract}: Ordered FePdCu nanoisland arrays were formed by annealing at 600$^{\circ}$C, which caused solid state dewetting of [Cu/Fe/Pd] multilayers deposited on self-assembled SiO$_{2}$ nanospheres with size of 100~nm. 
     A~single FePdCu island was formed on the top of each SiO$_{2}$ nanosphere. 
     The structure of the obtained system was studied by x-ray diffraction (XRD), while its magnetic properties by SQUID magnetometry. 
     Partially ordered L1$_{0}$ alloy appeared in the annealed films, leading to magnetic hardening of the material. 
     The paper presents the influence of the patterning on the system properties. 
     It is shown that templated dewetting is a~method providing nanoislands with well controlled sizes and positions. 
     The role of copper admixture in controlling the structural and magnetic properties is also discussed.
     
     \vspace{2ex}
     DOI: 10.1088/0957-4484/26/42/425301
     
     \vspace{2ex}
     Keywords: arrays, magnetic alloys, patterning, dewetting, self-assembly, thin films
     
    \vspace{3ex}
  }
  \end{@twocolumnfalse}
]

\section{Introduction}

One of the driving forces behind the development of innovative patterned magnetic materials is the magnetic recording industry. 
A promising idea that could lead to an increase of recording density is the use of Bit Patterned Media (BPM) concept,\cite{1,2} according to which one bit of information is stored in an individual magnetically isolated nanostructure. 
For these applications, the fabrication method must be low-cost and efficient enough to produce regular arrangements of nanostructures over large areas.\cite{3} 
Currently available nanopatterning methods, such as electron-beam lithography or focused ion beam lithography, allow to control the size and shape of nanostructures but are too expensive for practical applications.\cite{4}

One of the promising way of nanomagnet arrays fabrication is to use  porous alumina templates as substrates.\cite{5} 
The evaporation, sputtering or electrodeposition\cite{6} of thin films on such substrates consisting of regularly distributed nanometer sized artificial defects results in hardening of their magnetic properties.\cite{6,7} 
This effect appears as local anisotropy dependent on size, shape and arrangement of pores which influences the coercivity,  remanence, domain distribution and magnetization reversals.\cite{7,8,9} 
This technique allows to study magnetic phenomena for a variety of templates; the distance between the pores can be changed from 50 to 500~nm, and their diameter from 6~nm to 200~nm.\cite{5,6} 
However, the low degree of long-range order and a typical domain size being approximately a~few microns are the main disadvantages of this method. 
The arrangement of the nanostructures can be improved by using surface prepatterning with corrugations or imprinting stamps,\cite{6,10} but it is still difficult to obtain macroscopic order on the surface of square centimeters. 

As an alternative, nanoparticle arrays can also be made by the spontaneous solid state dewetting (SSD) of a metal film on a prepatterned substrate.\cite{11} 
SSD is induced by the minimization of energies associated with the interfaces present in the system, and can result from thermal processes, such as annealing, or laser or ion beam irradiation. 
It proceeds by surface diffusion that usually starts at film edge, and due to a large gradient of edge curvature moves atoms from the edge to the middle of the film leading to retraction and thickening of edges and creation of holes in the next step. 
It leads to separation of the thin film into nanoparticles, whose position is determined by the morphology of the substrate. 
Substrates with pits\cite{11,12,13} or ripples\cite{14,15,16} are used most often; however, this effect can be accelerated by using the patterned substrate with large curvature of surface, such as self-assembled arrays of nanoparticles. 
Their advantage is low cost, high simplicity of manufacturing, and large scale order over areas bigger than 1~cm$^{2}$. 
An additional benefit is high scalability. 
Commercial monodisperse suspensions of nanospheres are available in sizes ranging from 20~nm to several micrometers.

The templated solid-state dewetting has already been successfully used to pattern pure metal layers such as Co, Ni, Au or Ag.\cite{11,12,13,14,17} 
However, the technique has not been yet extended to complex metallic systems, such as alloys or ternary solid solutions. Currently, the research concerning patterned materials for magnetic recording is focused on finding a suitable magnetic material, allowing stable storage of one bit of information on structures smaller than 100~nm. 
For this purpose, thin alloy films revealing L10 structure appear to have a great potential.\cite{18} 
FePd alloys are a~good example of such materials. 
The addition of an admixture (e.g. copper) to FePd alloy modifies its growth and induces changes in crystallographic structure.\cite{19,20} 
This in turn causes changes in magnetic properties. 
Tailoring of the macroscopic and local magnetic parameters of layers is therefore possible through a careful choice of appropriate admixture and its introduction into the system. Numerous experiments\cite{19,20,21,22} showed that copper seems to be very promising in tuning the magnetic properties but its role in the patterned systems has not been yet examined.

\begin{figure*}[t!]
\centering
\includegraphics[width=0.7\textwidth]{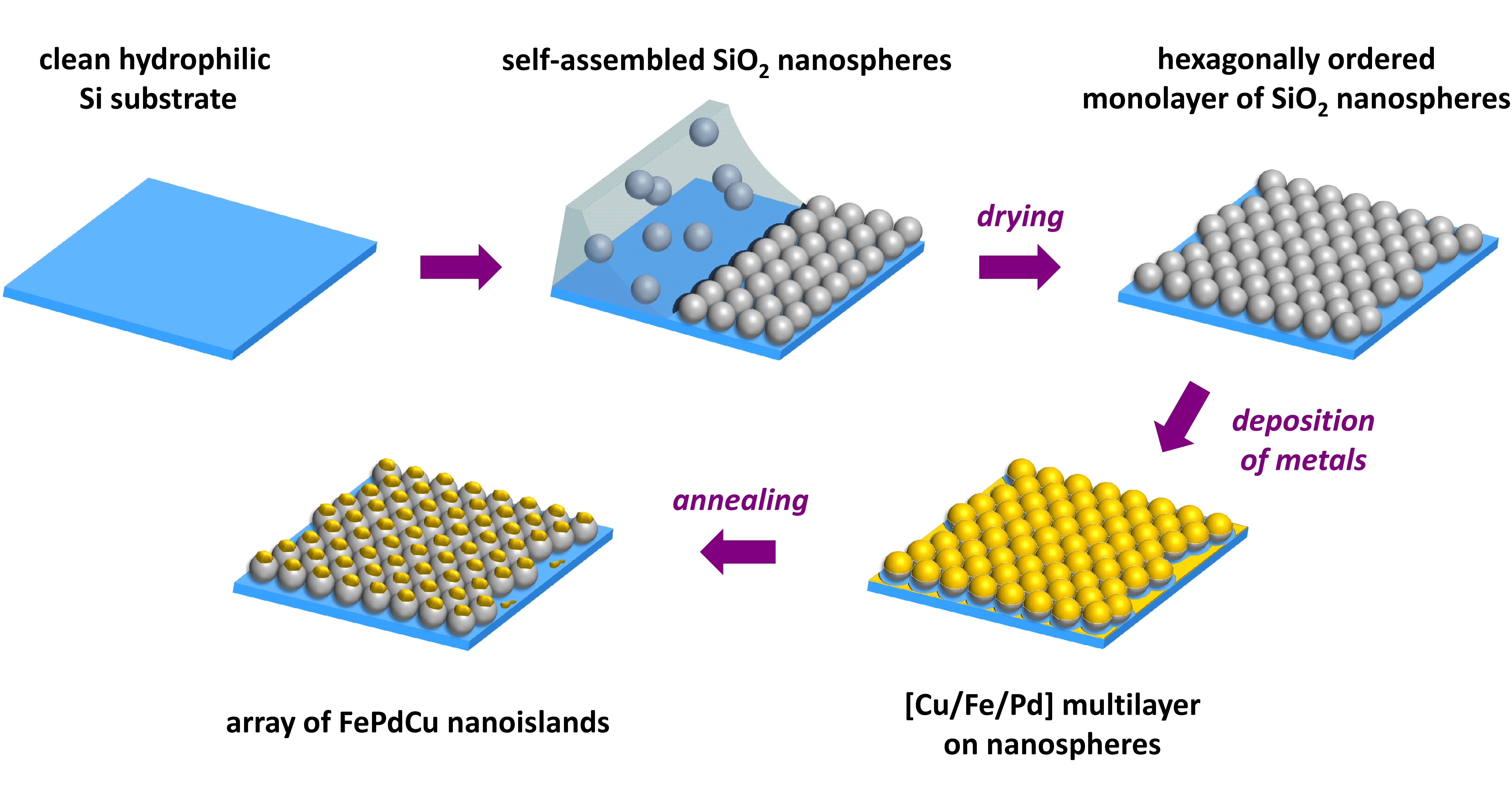}
\caption{General scheme of samples preparation.}
\label{Fig_1}
\end{figure*}
The main goal of the work is to determine structure and magnetic properties of the ordered FePdCu nanoisland arrays fabricated by the templated solid-state dewetting. 
Self-assembled monolayers of SiO$_{2}$ nanospheres were used as prepatterned substrates. 
The main objective of the studies is to investigate the influence of the patterning method and the role of admixture on the magnetism of the alloy system. 
The study shows simple one-step treatment leading to the creation of ordered arrays of FePdCu alloy.

\section{Experimental details}

General scheme of sample preparation is shown in Figure~\ref{Fig_1}. 
In order to produce prepatterned substrates, monolayers of 100~nm SiO$_{2}$ nanospheres were made using a method proposed by Micheletto et al.\cite{23} 
The basic idea of this method is a~deposition of a droplet of particle-water mixture onto a tilted, hydrophilic substrate. 
For this purpose, the Si wafers were washed in an ultrasonic cleaner in a~solution of hydrogen peroxide and ammonia, and then in a mixture of hydrogen peroxide and hydrochloric acid. 
In the last stage the substrates were rinsed with distilled water at room temperature, and then dried with nitrogen. 
Shortly after this preparation process, the droplet of the monodisperse SiO$_{2}$ nanoparticle suspension in water was placed onto the substrate. 
The solution was prepared by diluting commercially available 5\% (w/w) dispersion of amorphous silica spheres in water from Bang Laboratories. 
As reported by the manufacturer, SiO$_{2}$ nanospheres were nonporous and non-functionalized. 
A~hexagonal close packed particle array was formed during the evaporation of the solvent.

Then, the Si substrates with arrays of SiO$_{2}$ nanospheres were introduced into preparation chamber and the [Cu(t nm)/Fe(0.86~nm)/Pd(1.14 nm)]$_{\mathrm{x5}}$ multilayers were deposited on them by sequential thermal evaporation. 
The thickness of Fe and Pd layers was chosen as corresponding to 48:52 atomic stoichiometry, necessary to obtain the L1$_{0}$ structure.\cite{24} 
Copper was introduced into each Cu/Fe/Pd trilayer in an amount corresponding to the thickness $t$ of 0.1~nm or 0.2~nm, which resulted in two contents of copper in the system: approx. 5 at.\% and 10 at.\%. 
Layers were deposited at room temperature, with the evaporation rates of 0.5~nm/min for Fe and Pd, and 0.3~nm/min for Cu. 
Source materials used for the evaporation were of 99.9\% or higher purity. 
The working pressure was approximately 10$^{-9}$~mbar and the film thickness was controlled during evaporation with a~quartz thickness monitor.

The same multilayer systems were deposited also on flat Si wafers covered by 100~nm thick SiO$_{2}$ layer, in order to produce reference flat samples. 
The substrates were purchased from CrysTec and were produced by thermally oxidization of polished Si(001) wafers. 
Their rms roughness was approximately 0.5~nm, as confirmed by x-ray reflectivity studies. 
Prior to deposition of metals, the substrates were ultrasonically cleaned in acetone and ethanol and then rinsed in de-ionized water. 
Since the surface of the nanospheres was not functionalized, we assumed that the surface groups on the flat substrates are the same as on silica nanospheres. 
Therefore, from the chemical point of view, properties of both systems are similar. 
The parameters of the deposition were the same as in the case of patterned systems. 

The samples were then annealed in vacuum chamber with base pressure of 10$^{-7}$~mbar. 
Heating rate of 10$^{\circ}$C/min was applied to heat up the samples to the temperature of 600$^{\circ}$C. 
The annealing time at the final temperature was set to 1~min. 
After this time the heating power was turned off and the samples were left under vacuum until they reached room temperature.

Both before and after the annealing the samples were examined by x-ray diffraction (XRD) in Bragg-Brentano geometry. 
The experiments were performed with a two-circle diffractometer (Panalytical X’Pert Pro) using Cu $K_{\alpha}$ line at 1.54~\AA. 
The XRD patterns were measured in the standard $\Theta$-$2\Theta$ geometry for $2\Theta$ angles ranging from 20~deg to 90~deg, with the instrumental step of 0.05~deg. 
The signal was collected by solid state stripe detector with a graphite monochromator.

The magnetic measurements were performed using a SQUID magnetometer with a maximal magnetic field of $\pm$70~kOe. 
All measurements were carried out at room temperature in out-of-plane and in-plane geometry of the applied magnetic field. 
After the measurement, the magnetic background from the substrate was subtracted from the measured signal. 
Then, a~correction for the magnetic moment associated with the shape of the sample was performed, according to the Quantum Design application note.\cite{25}

\section{[Cu/Fe/Pd] multilayers}

The chemical composition of the as-deposited samples was checked by Rutherford backscattering spectrometry, which confirmed the nominal composition of the films. 
The layered structure of the samples was investigated using x-ray reflectivity (XRR), which corroborated nominal thicknesses of the layers. 
XRR results also showed that the average roughness of the layers had a~value of about 0.6~nm. 
This indicates a~significant intermixing between them during the deposition process.

\begin{figure}[t]
\centering
\includegraphics[width=0.44\textwidth]{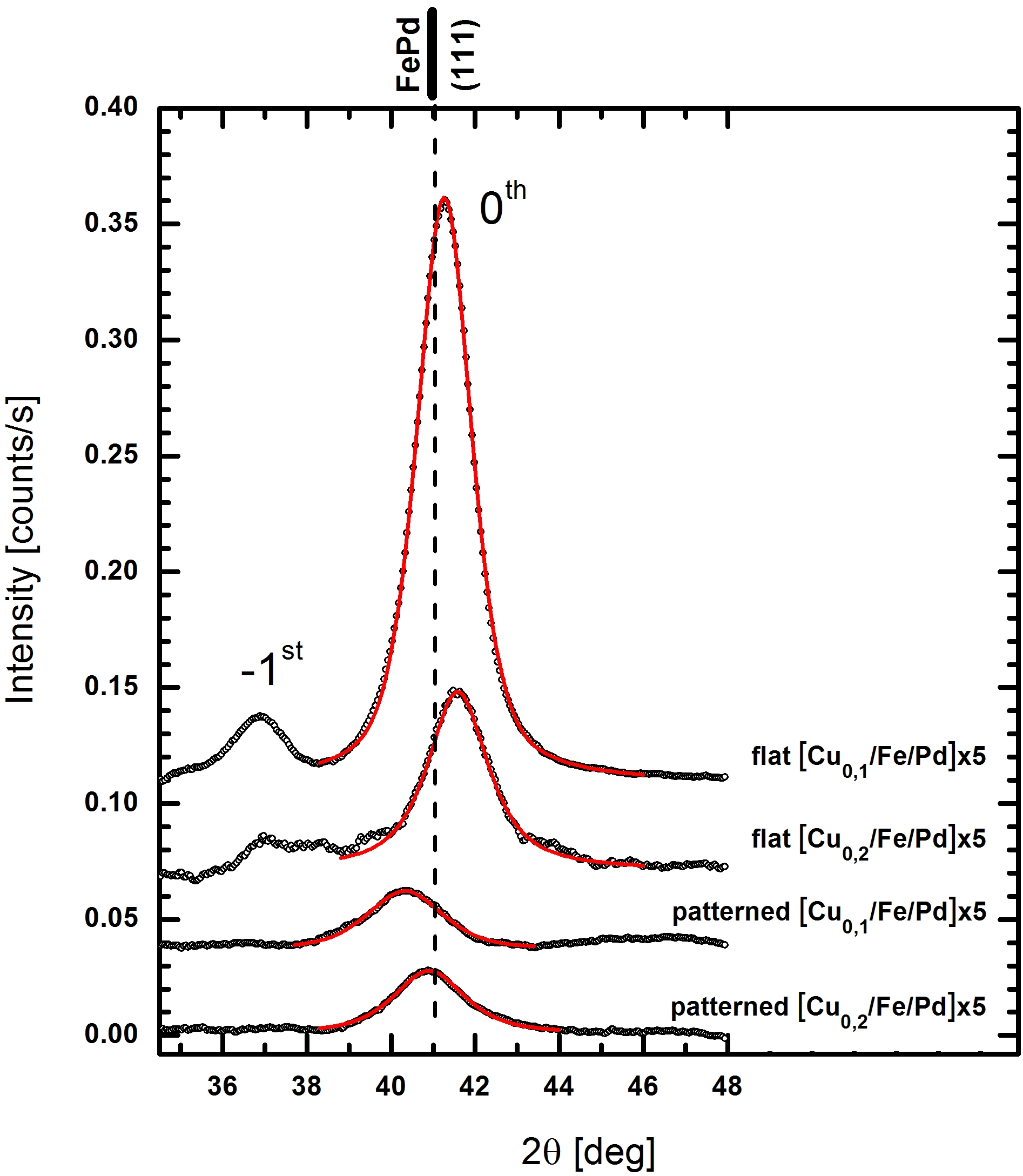}
\caption{XRD patterns of the [Cu/Fe/Pd] multilayers deposited on the flat Si/SiO$_{2}$ substrates, as well as on the SiO$_{2}$ nanospheres with the size of 100~nm. Vertical offsets were used for clarity. The position of the (111) reflection for the bulk FePd alloy is indicated.}
\label{Fig_2}
\vspace{-0.2cm}
\end{figure}

In x-ray diffraction patterns, shown in Figure~\ref{Fig_2}, only one broad peak at $2\Theta \approx 41$~deg was observed together with a~small first order satellite peak appearing for the flat layers at an angle $2\Theta$ of approximately 37~deg. 
It has been previously reported\cite{26,27,28,29,30} that such satellite peaks are related to periodically layered structures. 
The angular distance between main and satellite peak is approx. 4.5~deg and corresponds to the thickness of the [Cu/Fe/Pd] trilayer calculated according to ref~[31]. 
The position of the main peak is close to the (111) peak of bulk FePd alloy, and may come from the disordered FePdCu alloy created due to interlayer diffusion during the deposition process. 
The peak position shifts to the larger angles with the increasing Cu layer thickness, which indicates a decrease of distance between the (111) planes. 
A~similar shift of the (111) peak due to the addition of copper have been reported by Polit et al.\cite{32} 
Detailed studies with extended x-ray absorption fine structure (EXAFS) indicate that the copper admixture changes the radii of the first coordination shells, which results in changes of the lattice constants.\cite{33}

Reflexes from pure Pd, Fe or Cu did not appear in the spectrum, which confirms the substantial intermixing of the metals already during deposition. 
Coherence lengths, corresponding to the size of the grains, were estimated using the Scherrer formula. 
Their values for the flat systems amount to 5~nm, while for the patterned systems they are smaller and amount to approximately 4~nm.

\begin{figure*}[t]
\centering
\includegraphics[width=0.8\textwidth]{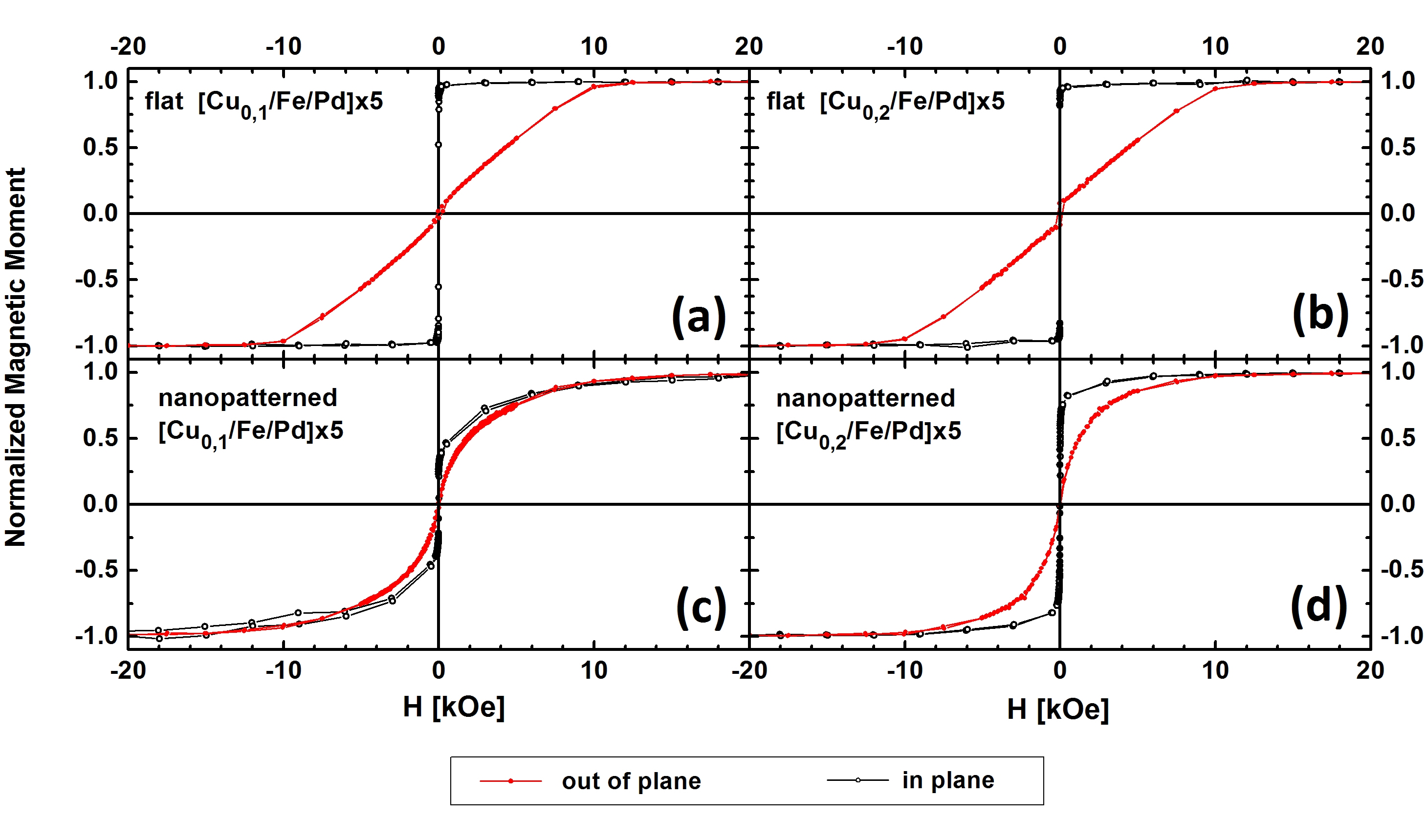}
\caption{Magnetic hysteresis loops measured in the out-of-plane and in-plane geometry for [Cu/Fe/Pd] multilayers deposited on (a,b) the flat Si/SiO$_{2}$ substrate, (c,d) arrays of the SiO$_{2}$ nanospheres with the size of 100~nm.}
\label{Fig_3}
\end{figure*}

Figure~\ref{Fig_3} shows magnetic hysteresis loops obtained at room temperature, while the values of coercive fields are presented in Table~\ref{Table_1}. 
\begin{table}[t]
\centering
 \begin{tabular}{l c c} 
 \hline \hline
 \textbf{Sample} & \multicolumn{2}{c}{\boldmath$H_{\mathrm{c}}$ [Oe]} \\ \hline
                 & \textbf{\textit{in-plane}} & \textbf{\textit{in-plane}} \\ \cline{2-3}
flat  [Cu$_{0.1}$/Fe/Pd]x5 & 7 $\pm$ 2 & 124 $\pm$ 37 \\
flat  [Cu$_{0.2}$/Fe/Pd]x5 & 3 $\pm$ 3 & 145 $\pm$ 20 \\
patterned  [Cu$_{0.1}$/Fe/Pd]x5 & 9 $\pm$ 2 & 44 $\pm$ 9 \\
patterned  [Cu$_{0.2}$/Fe/Pd]x5 & 6 $\pm$ 1 & 24 $\pm$ 3 \\ \hline\hline
 \end{tabular}
\caption{The in-plane and out-of-plane values of coercive field $H_{\mathrm{c}}$, derived from SQUID measurements for the flat and patterned [Cu/Fe/Pd] multilayers.}
 \label{Table_1}  
 \vspace{-0.2cm}
\end{table}
For the multilayers deposited on planar Si/SiO$_{2}$ substrates (Figures~\ref{Fig_3}a and b), the clear difference is observed between the shape of the loops measured for the field directed in-plane and out-of-plane with respect to the sample surface. 
For the field parallel to the sample surface, the saturation magnetization is reached at field approx. 10~kOe lower than for the out-of-plane geometry indicating that the easy axis of magnetization is in the sample plane. 
Such effect is related to the shape anisotropy, which dominates for the [Fe/Pd/Cu] multilayers grown at room temperature. 
The in-plane coercivity is negligibly small. Only for the perpendicular direction of the external magnetic field the coercive field is non-zero and amounts to approximately 100~Oe. 

The magnetic hysteresis loops for [Fe/Pd/Cu] multilayers deposited on the SiO$_{2}$ nanospheres exhibit different features. 
Magnetic easy axis is still parallel to the substrate plane; however, the hysteresis loops for in- and out-of-plane geometry are closer to each other, becoming more isotropic. 
Such behaviour is directly caused by the curved surface of the SiO$_{2}$ nanospheres. 
Shape anisotropy forces the magnetization to be in a~film plane, but because of the spherical shape of films, both components of the magnetization (parallel and perpendicular) appear. 
This effect is stronger for thinner multilayers with smaller amount of copper.

The values of the magnetic moment of the flat samples, calculated assuming that it is located only on the iron atoms, amount to approximately 2.45~${\mu_{\mathrm{B}}}$/atom. 
They are bigger than for pure bcc iron (2.2~${\mu_{\mathrm{B}}}$/atom), but on the other hand, smaller than the value for the bulk ordered FePd alloy (2.8~${\mu_{\mathrm{B}}}$/atom).\cite{34} 
This suggests that in these samples a part of iron atoms formed a~poorly ordered FePd alloy already during the deposition process. 
Approximately three times lower values of Bohr magnetons per iron atom are observed for the patterned samples. 
The difference may result from the complex curvature of film surface connected with the reduction of grain size in patterned multilayers. 
According to [35], small crystallographic grains with a radius smaller than 5~nm show typical superparamagnetic features with the saturation magnetization substantially smaller than those observed for the bulk. 
In our case, the patterned samples have complex developed surface, and the formation of small grains is more likely, particularly at the edges of the SiO$_{2}$ nanospheres. 
XRD measurements confirmed that the patterned samples have approximately 30\% smaller coherence length than the flat samples. 
Similar effect was already observed for other nanosystems including nanoparticles, nanocrystalline materials, and thin films.\cite{36,37,38,39}

\section{FePdCu alloys}

\begin{figure}[t]
\centering
\includegraphics[width=0.44\textwidth]{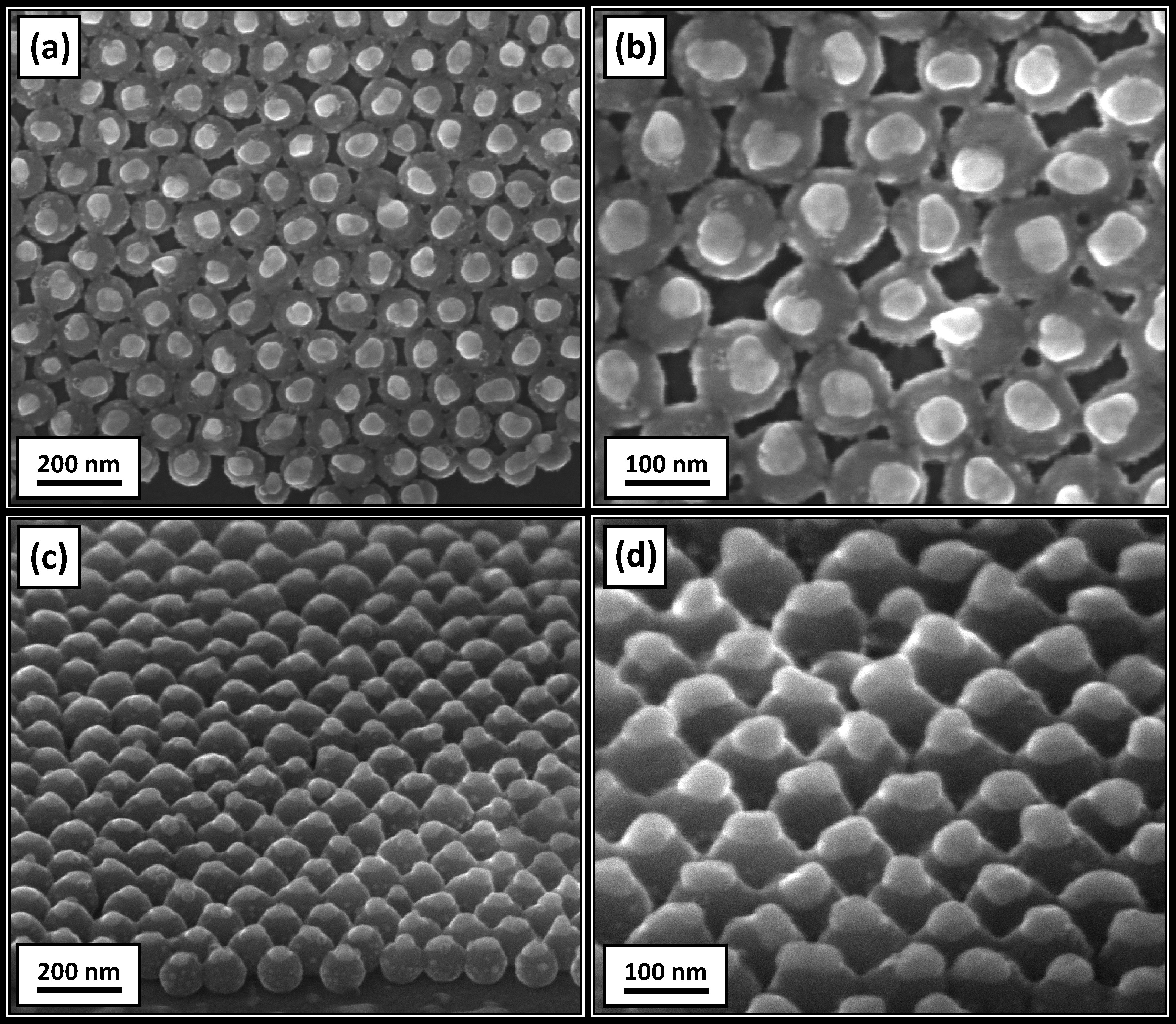}
\caption{Arrays of the nanospheres with the size of 100~nm covered by the FePdCu alloy with 5 at.\% of the Cu admixture. Pictures (a) and (b) were collected with the incidence angle of 0~deg while (c) and (d) with the incidence angle of 45 deg. Alloy islands appear bright due to the charging effects.}
\label{Fig_4}
\end{figure}

\begin{figure}[h]
\centering
\includegraphics[width=0.47\textwidth]{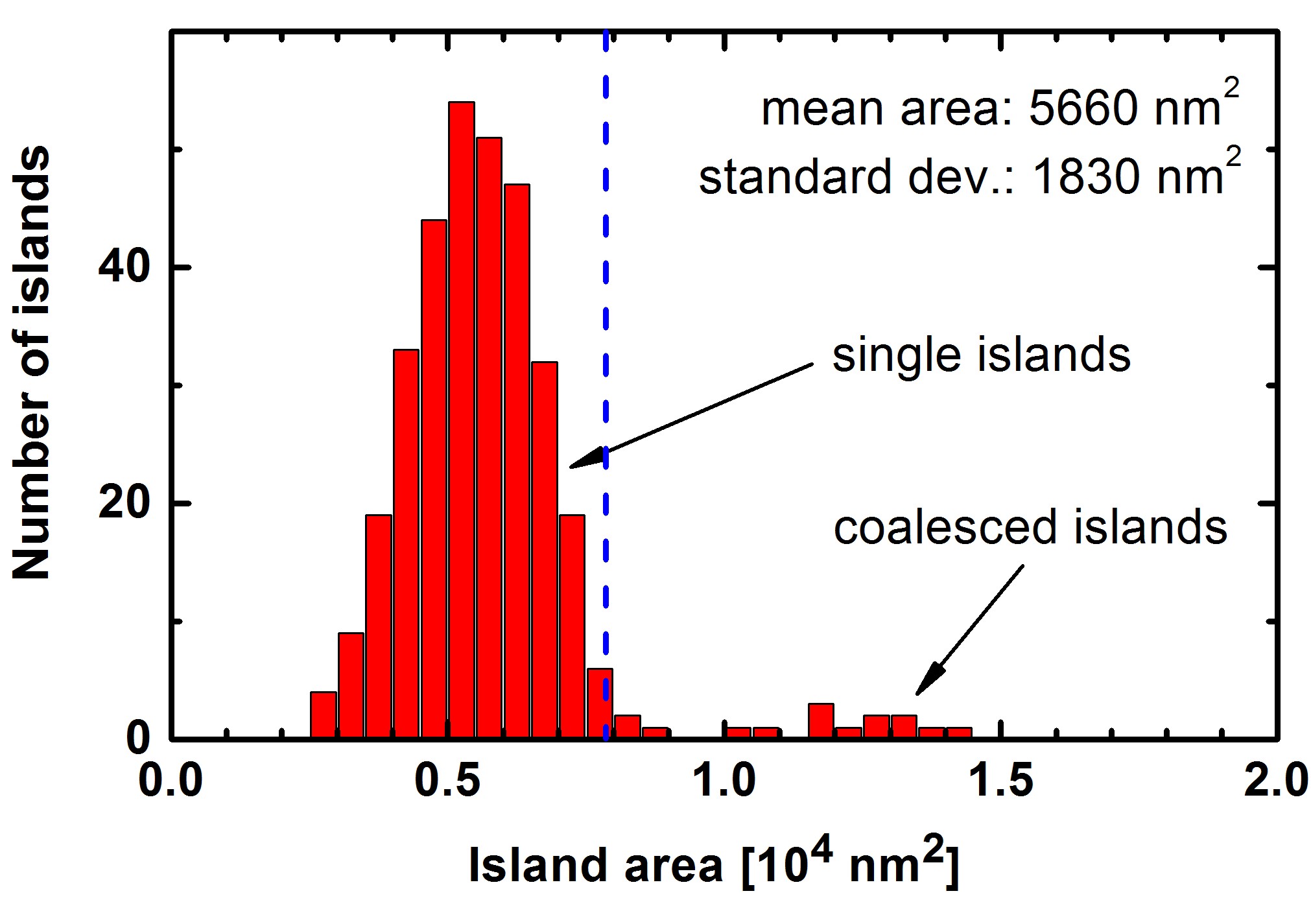}
\caption{Histogram of the island size distributions obtained from SEM images for the agglomerated alloy on the SiO$_{2}$ nanospheres with the size of 100~nm. The cross section area for the single SiO$_{2}$ 100 nm nanoparticle was indicated by the blue dashed line.}
\label{Fig_5}
\end{figure}

\begin{figure}[h]
\centering
\includegraphics[width=0.47\textwidth]{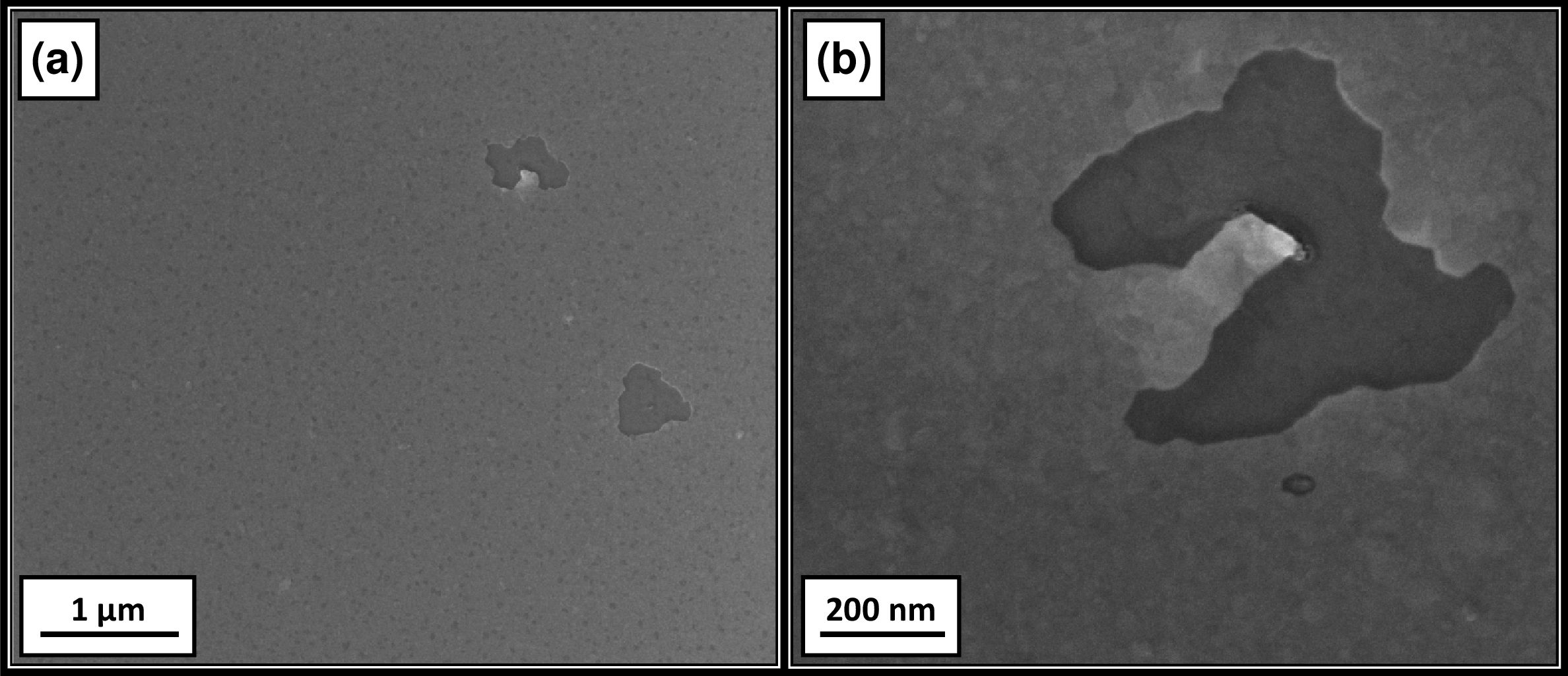}
\caption{Flat Si/SiO$_{2}$ substrates covered by the FePdCu alloy with 5 at.\% of Cu admixture. Picture (a) shows the selected location on the sample, with traces of delamination and dewetting. Picture (b) presents the details of one selected void.}
\label{Fig_6}
\end{figure}

After the annealing the layers deposited on flat substrates remain continuous, while those on SiO$_{2}$ nanospheres split up into isolated islands (see Figure~\ref{Fig_4}). 
The island arrangement reflects the hexagonal order of the SiO$_{2}$ nanospheres array, and a~single island of material is located on the top of each nanosphere. 
The mean diameter of the islands is 45~nm, and their size distribution is shown in Figure~\ref{Fig_5}.
Approximately 5\% of the islands has been joined together, due to coalescence process. 
The reference flat FePdCu samples exhibited only small morphological changes in the form of rarely appearing voids, which are showed in Figure~\ref{Fig_6}.

All of the morphological changes can be explained by the phenomenon of solid-state dewetting, which is a~thermally activated process resulting in a~reduction of the free energy of the system. 
The driving force for this process is the reduction of the film-substrate interfacial energy, film-surface interfacial energy and stress within the film, but usually the last contribution is negligible.\cite{40,41} 
Additionally, Petersen et al\cite{14} showed that solid-state dewetting takes place at lower temperature on a~prepatterned substrate than for the flat systems. 
Due to the additional driving force of curvature induced diffusion, the system needs lower thermal energy to create initial voids leading to formation of islands and further coalescence. 
Thus, in the case of the alloy deposited on the curved surface of nanoparticles, the dewetting process occurred already at the temperature of 600$^{\circ}$C, well below the melting point of the alloy, while for flat samples the same annealing conditions were not sufficient to create islands. 
The process initiated creation of rarely distributed voids, which did not change significantly the morphology of the layer.

\subsection{Structure}

\begin{figure}[h]
\centering
\includegraphics[width=0.44\textwidth]{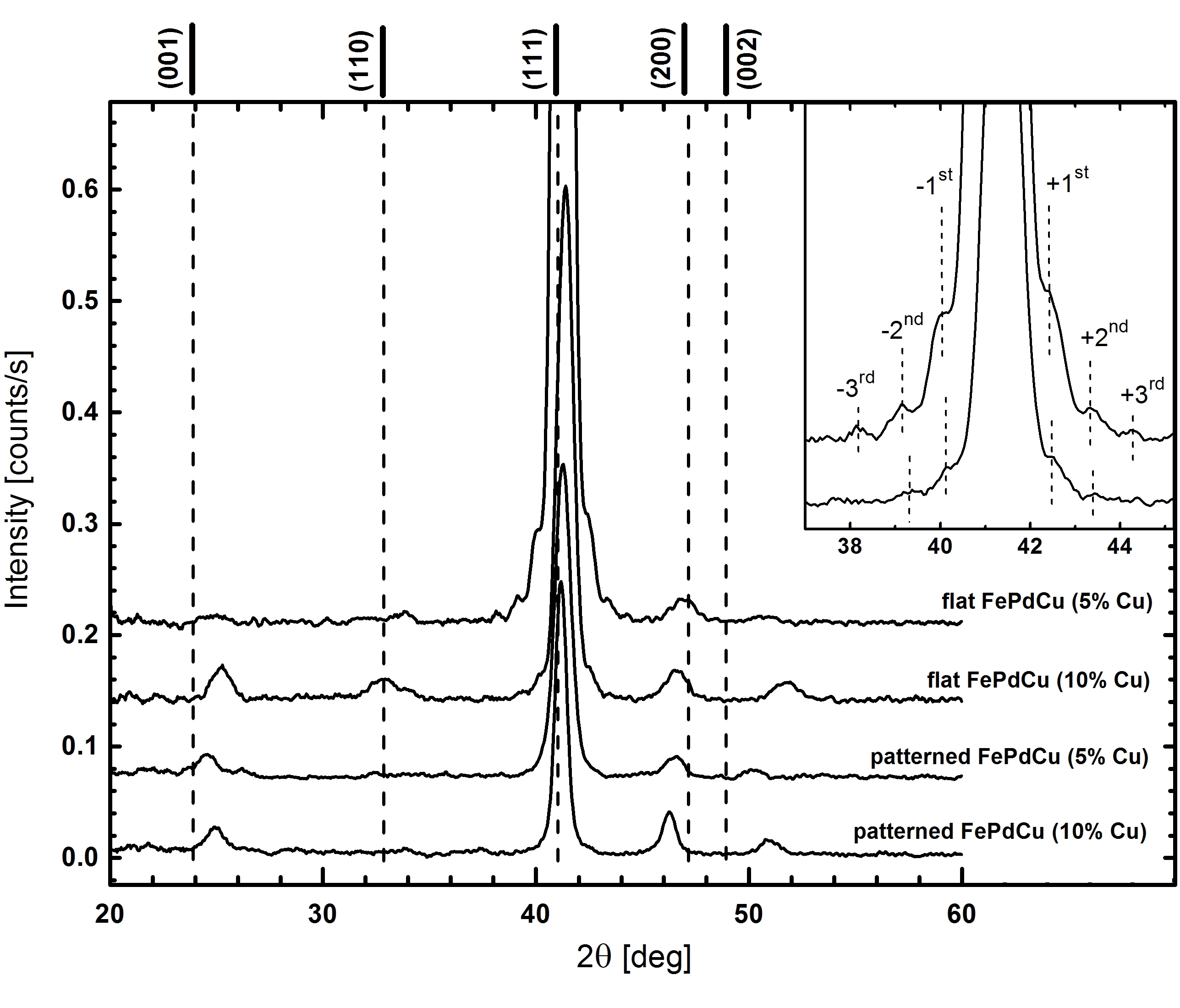}
\caption{XRD patterns of the FePdCu alloys deposited on the flat Si/SiO$_{2}$ substrates, as well as on the SiO$_{2}$ nanospheres with the size of 100~nm. The vertical offsets were used for clarity. The positions of reflections for the bulk FePd alloy are indicated. The inset shows a~magnification of the patterns for flat systems in the vicinity of the (111) peak. The positions of satellite peaks are indicated.}
\label{Fig_7}
\end{figure}

The x-ray diffraction patterns for the FePdCu alloys deposited on the flat and prepatterned substrates are shown in Figure~\ref{Fig_7}. 
The positions of the reflections for the bulk L1$_{0}$-ordered FePd alloy are marked with the vertical dashed lines. 
The comparison of the peak positions allow identification of the observed reflections as (001), (110), (111), (200) and (002) reflections belonging to the L1$_{0}$ crystallographic structure of FePdCu alloy. 
The presence of the (001) reflection, prohibited in the face-centered cubic disordered alloy, indicates that the unit cell of the alloy is tetragonal. 
This is also confirmed by the separation of the (200) and (002) peaks. 
This confirms the presence of the distortion in the crystallographic cell characteristic for the L1$_{0}$ structure. 
It demonstrates that the transformation from the initial multilayer resulted in fct-ordered FePdCu alloy formation. 

All XRD patterns show a strong (111) peak related to the [111]-oriented crystallites. 
In the case of flat systems, small satellite peaks surrounding the main diffraction peak can be observed. 
They are an indication of high homogeneity and flatness of the layers and originate from the interference between x-ray radiation scattered from the bottom and the top surfaces of the film. 
The thicknesses of the film, calculated from the distance between position of the peaks, are equal to 9.6~nm $\pm$ 0.5~nm and 9.8 $\pm$ 0.7~nm for flat FePdCu alloys with 5 at.\% and 10 at.\% of Cu, respectively.

The line profiles were fitted with the pseudo-Voigt function. 
The crystallographic lattice constants and the $c/a$ ratios are summarized in Table~\ref{Table_2}. 
\begin{table}[t]
\caption{Summary of $a$ and $c$ lattice parameters, the $c/a$ axial ratios, and the long-range order parameters $S$ obtained from the XRD patterns for the FePdCu alloys deposited on the flat as well as on the nanopatterned substrates.}
\vspace{0.4cm}
\centering
 \begin{tabular}{c c c} 
 \hline \hline
           & \multicolumn{2}{c}{\textbf{FePdCu (5 at.\% Cu)}} \\ \cline{2-3}
           & ~~~~~~\textbf{flat}~~~~~~     & ~~~~~~\textbf{patterned}~~~~~~ \\ \hline
\rule[3ex]{0.1ex}{0ex} $a$ [\AA]  & 3.869 $\pm$ 0.012 & 3.903 $\pm$ 0.010 \\
\rule[2ex]{0.1ex}{0ex} $c$ [\AA]  & 3.589 $\pm$ 0.011 & 3.627 $\pm$ 0.010 \\
\rule[2ex]{0.1ex}{0ex} $c/a$      & 0.928 $\pm$ 0.006 & 0.929 $\pm$ 0.005 \\
\rule[2ex]{0.1ex}{0ex} $S$        & 0.47              & 0.79 \\ \hline\hline
           & \multicolumn{2}{c}{\textbf{FePdCu (10 at.\% Cu)}} \\ \cline{2-3}
           & \textbf{flat}     & \textbf{patterned} \\ \hline
\rule[3ex]{0.1ex}{0ex} $a$ [\AA]  & 3.895 $\pm$ 0.011 & 3.922 $\pm$ 0.009 \\
\rule[2ex]{0.1ex}{0ex} $c$ [\AA]  & 3.529 $\pm$ 0.010 & 3.573 $\pm$ 0.008 \\
\rule[2ex]{0.1ex}{0ex} $c/a$      & 0.906 $\pm$ 0.006 & 0.911 $\pm$ 0.005 \\
\rule[2ex]{0.1ex}{0ex} $S$        & 0.75              & 0.77 \\ \hline\hline
 \end{tabular}
 \label{Table_2}  
\end{table}
For all samples the $c/a$ ratios are smaller than the value for the pure FePd alloy (0.966), and additionally are influenced by concentration of copper admixture. 
A~similar change of the lattice parameter values in the FePdCu system was already reported in [19,32]. 
The analysis indicates also that patterning does not influence the $c/a$ ratio.

The values of the long-range order parameter $S$ were calculated using a~method from~[42] and are summarized in the last lines of Table~\ref{Table_2}. 
In all cases the values of the order parameter are smaller than for the FePd alloys deposited on MgO at the elevated temperature by molecular beam epitaxy.\cite{43,44,45} 
However, they are comparable and even greater than the $S$ values for the FePd alloys deposited on MgO at room temperature\cite{46} which were equal to 0.58. 
It is in agreement with results of [47] which show that the tensile stress present in polycrystalline multilayers is more important to obtain high degree of the long range order in FePd alloy. 

In addition to the analysis of the lattice constants, the coherence lengths were calculated from Scherrer equation. 
Their values are summarized in Table~\ref{Table_3}. 
\begin{table}[h]
\caption{Summary of the coherence lengths $L$ calculated for (001), (111), (200) and (002) crystallographic directions, obtained from the XRD patterns for the FePdCu alloys deposited on the flat as well as on the nanopatterned substrates. For each sample a~mean coherence length was determined as an arithmetic mean of the values for four crystallographic directions.}
\vspace{0.4cm}
\centering
 \begin{tabular}{c c c} 
 \hline \hline
           & \multicolumn{2}{c}{\textbf{FePdCu (5 at.\% Cu)}} \\ \cline{2-3}
           & ~~~~~~\textbf{flat}~~~~~~     & ~~~~~~\textbf{patterned}~~~~~~ \\ \hline
\rule[2ex]{0.1ex}{0ex} $L_{001}$ [nm]  & 5.1 $\pm$ 1.5  & 6.0 $\pm$ 0.9 \\
$L_{111}$ [nm]  & 10.7 $\pm$ 0.7 & 10.4 $\pm$ 0.6 \\
$L_{200}$ [nm]  & 7.3 $\pm$ 0.7  & 9.3 $\pm$ 0.8 \\
$L_{002}$ [nm]  & 6.7 $\pm$ 1.1  & 9.4 $\pm$ 1.2 \\ 
$\bar{L}$ [nm]  & 7.5            & 8.8 \\ \hline\hline
           & \multicolumn{2}{c}{\textbf{FePdCu (10 at.\% Cu)}} \\ \cline{2-3}
           & \textbf{flat}     & \textbf{patterned} \\ \hline
\rule[2ex]{0.1ex}{0ex} $L_{001}$ [nm]  & 8.3 $\pm$ 0.8  & 9.1 $\pm$ 0.7  \\
$L_{111}$ [nm]  & 10.9 $\pm$ 0.7 & 14.4 $\pm$ 1.2 \\
$L_{200}$ [nm]  & 7.9 $\pm$ 1.2  & 12.0 $\pm$ 1.0 \\
$L_{002}$ [nm]  & 7.6 $\pm$ 0.6  & 9.1 $\pm$ 0.7 \\ 
$\bar{L}$ [nm]  & 8.7            & 11.2 \\ \hline\hline
 \end{tabular}
 \label{Table_3} 
 \vspace{-0.2cm}
\end{table}
The calculations were done for the (001), (111), (002) and (002) reflections providing information about the crystallite sizes along the [001], [100], and [111] crystallographic directions. 
In all cases the values for the [111] direction were the highest, which indicates that the growth along this crystallographic direction is the easiest. 
An increase of the mean coherence length with increasing copper content can be seen when comparing the samples with different copper content, which may indicate that copper facilitates the diffusion of atoms at elevated temperature and accelerates the growth of the crystallites.

The increase of the coherence lengths for patterned samples in comparison with the flat systems can be also noticed. 
In order to explain this observation, it should be noted that the XRD in Bragg-Brentano geometry is sensitive only to the coherence length perpendicular to the sample plane. 
Flat systems did not significantly change their morphology after the annealing, which  means that their thickness also remained constant at about 11~nm. 
This determines the maximum coherence length for such systems also at 11~nm. 
On the contrary, the patterned systems have created islands, whose thickness is greater than the original thickness of the multilayer. 
This offers the space for further grain growth in the direction perpendicular to the sample surface, which is observed as an increase in the average coherence length.

\vspace{0.3cm}
\subsection{Magnetism}

\begin{figure}[b]
\centering
\includegraphics[width=0.49\textwidth]{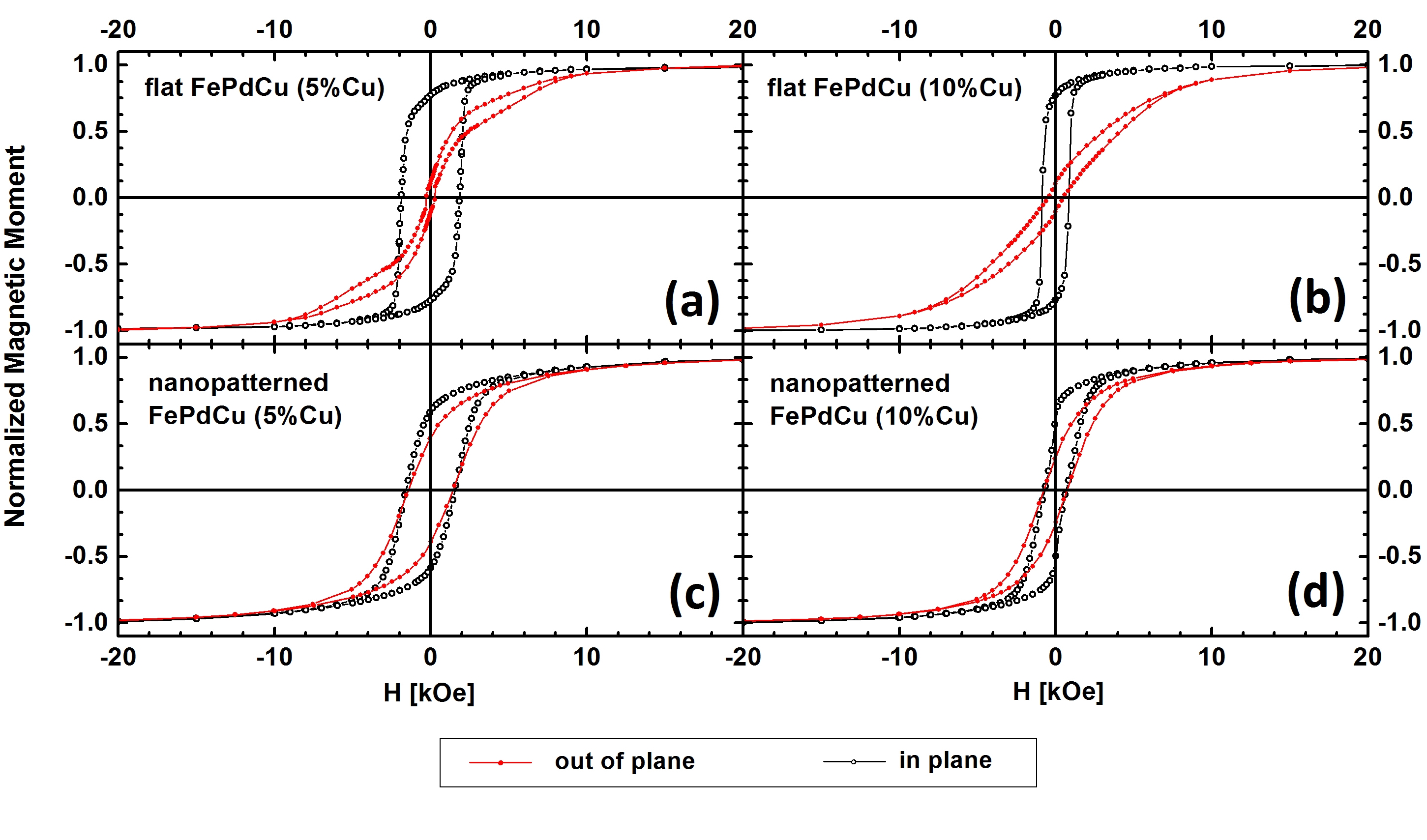}
\caption{Magnetic hysteresis loops measured in the out-of-plane and the in-plane geometries for the FePdCu alloys deposited on (a,b) the flat Si/SiO$_{2}$ substrate, (c,d) arrays of the SiO$_{2}$ nanospheres with the size of 100~nm.}
\label{Fig_8}
\vspace{0.2cm}
\end{figure}

Figure~\ref{Fig_8} shows magnetic hysteresis loops obtained at room temperature for the alloys with two different concentration of copper deposited on the flat substrates as well as on arrays of SiO$_{2}$ nanospheres. 
The values of coercive field are presented in Table~\ref{Table_4}. 
\begin{table}[t]
\caption{The in-plane and out-of-plane values of the coercive field $H_{\mathrm{c}}$, derived from SQUID measurements for the flat and patterned FePdCu alloys.}
\vspace{0.4cm}
\centering
 \begin{tabular}{l c c} 
 \hline \hline
 \textbf{Sample} & \multicolumn{2}{c}{\boldmath$H_{\mathrm{c}}$ [Oe]}  \\  \cline{2-3}
                 & \textbf{\textit{in-plane}} & \textbf{\textit{in-plane}} \\ \hline
flat  FePdCu (5 at.\% Cu) & 1861 $\pm$ 51 & 292 $\pm$ 32 \\
flat  FePdCu (10 at.\% Cu) & 850 $\pm$ 31 & 489 $\pm$ 25 \\
patterned  FePdCu (5 at.\% Cu) & 1551 $\pm$ 33 & 1412 $\pm$ 55 \\
patterned  FePdCu (10 at.\% Cu) & 724 $\pm$ 33 & 746 $\pm$ 29 \\ \hline\hline
 \end{tabular}
 \label{Table_4} 
 \vspace{-0.4cm}
\end{table}

After thermal processing, the character of hysteresis loops changed when compared to multilayers, and their features depend on the copper content. 
These changes are most apparent for the coercivity, which increased substantially by a~factor of 10 for the out-of-plane geometry and by a~factor of 100 for the in-plane geometry. 
The loops shape indicates that for all annealed samples the reversal of magnetization is easier for the external field directed along the plane of the sample. 
It is particularly clear for the flat samples. 
More isotropic magnetic behaviour is observed for the alloy on SiO$_{2}$ nanospheres. 
In this case the loop shapes and the coercive fields for the two measuring geometries are very similar. 
It is an indication that there is no preferential orientation of the magnetic easy axis. 
Such behaviour is directly related to the shape of islands, which is similar to an irregular ellipsoid. 
For such case magnetic shape anisotropy is not as strong as for the planar layers, which creates the possibility of nearly isotropic distribution of magnetic easy axes. 

The coercive field values for the patterned samples correspond to the values observed for the FePd nanoparticles with the similar size as island size determined from our SEM images.\cite{48,49} 
An increase of the Cu content results in the reduction of the coercivity. 
It correlates with a~small increase of the average coherence length with increasing copper content reported by the XRD measurements. 
It may indicate that copper favours the growth of the crystallites and lowers the amount of grain walls and defects serving as pinning sites, which leads to the lower coercivity. 
On the other hand, according to Naganuma\cite{20}, copper can also reduce magnetic coupling between the domains, which also could contribute to a~decrease of the coercive field value. 
However, a~more detailed investigation of the reversal behaviour of the alloy is required to verify these two proposed explanations of the coercivity decrease.

The values of Bohr magnetons per iron atom for the flat samples correspond well to the values reported by Shi\cite{34} for the ordered FePd alloy. 
Two times lower values of the magnetic moment per iron atom observed for patterned samples are likely related to the developed surface of the alloy, which contributes to the weakness of the magnetic coupling at the island boundaries. 
This phenomenon is accompanied by a decrease of the coercive field values.

\vspace{-0.3cm}
\section{Conclusions}

The use of solid-state dewetting to produce ordered arrays of ferromagnetic FePdCu alloy has been demonstrated on self-assembled matrices of SiO$_{2}$ nanospheres. 
This study showed that the dewetting templated by self-organized nanospheres can be successfully used to create L1$_{0}$ alloy nanoislands from [Cu/Fe/Pd] multilayers. 
One step annealing is sufficient to perform both patterning and alloying, leading to a system exhibiting the long-range order parameter amounting up to 0.79, which is comparable to the value for FePd alloys obtained by the epitaxial growth. 
Additionally, the patterning contributes to the increase of the grain size in comparison with the flat films. 
The patterning method influences also the magnetic properties. 
It causes a~decrease of the magnetic moment per iron atom and lowers the coercivity field value. 
The hysteresis behaviour is more isotropic for the nanopatterned samples than for the flat systems. 
It shows that the method can effectively induce the changes of the magnetic properties of L1$_{0}$ alloys and may be useful in functional devices that utilize the magnetic, optoelectrical, and plasmonic properties of ordered arrays of metal alloy nanoparticles.

\noindent
\vspace{0.1cm}

\textbf{Acknowledgements:} The work was partially supported by Polish National Science Center with Contract No. 2012/05/B/ST8/01818. 
The authors are grateful to the group of Prof. Manfred Albrecht from the University of Augsburg for performing of SQUID measurements.

\bibliographystyle{plainnat}
\vspace{0.1ex}
\begin{center}
 $\star$ $\star$ $\star$
\end{center}

\vspace{-1ex}

\setlength{\bibsep}{0pt}
\renewcommand{\bibnumfmt}[1]{$^{#1}$}

\end{document}